\documentclass[sigconf,natbib=false]{acmart}

\acmConference[Conference acronym 'XX]{Make sure to enter the correct
  conference title from your rights confirmation emai}{June 03--05,
  2024}{}

\acmISBN{978-1-4503-XXXX-X/18/06}

\begin{document}

\title{Metamorphic Evaluation of ChatGPT as a Recommender System}

\author{Madhurima Khirbat}
\affiliation{%
  \institution{RMIT University}
  \city{Melbourne}
  \country{Australia} }
  \email{madhurima.khirbat@student.rmit.edu.au}

\author{Yongli Ren}
\affiliation{%
  \institution{RMIT University}
  \city{Melbourne}
  \country{Australia}}
    \email{yongli.ren@rmit.edu.au}

\author{Pablo Castells}
\affiliation{%
  \institution{Universidad Autónoma de Madrid}
  \city{Madrid}
  \country{Spain}
}
\email{pablo.castells@uam.es}

\author{Mark Sanderson}
\affiliation{%
  \institution{RMIT University}
  \city{Melbourne}
  \country{Australia}
}
\email{mark.sanderson@rmit.edu.au}


\begin{abstract}
   With the rise of Large Language Models (LLMs) such as ChatGPT, researchers have been working on how to utilize the LLMs for better recommendations. 
  However, although LLMs exhibit black-box and probabilistic characteristics (meaning their internal working is not visible), the evaluation framework used for assessing these LLM-based recommender systems (RS) are the same as those used for traditional recommender systems.
  To address this gap, we introduce the metamorphic testing for the evaluation of GPT-based RS. 
  This testing technique involves defining of metamorphic relations (MRs) between the inputs and checking if the relationship has been satisfied in the outputs. 
  Specifically, we examined the MRs from both RS and LLMs perspectives, including rating multiplication/shifting in RS and adding spaces/randomness in the LLMs prompt via prompt perturbation. 
  Similarity metrics (e.g. Kendall $\tau$ and Ranking Biased Overlap(RBO)) are deployed to measure whether the relationship has been satisfied in the outputs of MRs. 
  The experiment results on MovieLens dataset with GPT3.5 show that lower similarity are obtained in terms of Kendall $\tau$ and RBO, which concludes that there is a need of a comprehensive evaluation of the LLM-based RS in addition to the existing evaluation metrics used for traditional recommender systems.
\end{abstract}





\maketitle

\section{Introduction}
The rise and advancements of Natural Language Processing (NLP) based systems involving Large Language Models (LLMs), such as Generative Pre-trained Transformer (GPT), BERT, LLaMa and many more have led to the possibility of automation of tasks. Recommender systems are no exception with researchers experimenting with LLM based Recommender Systems and incorporating LLMs into existing recommended systems, such as Collaborative Filtering and Matrix Factorization \cite{liu2023chatgpt, di_palma_evaluating_2023, dai_uncovering_2023, 10.1145/3523227.3546767}.
To evaluate the performance of LLM-based RS, existing research follows the same evaluation framework for traditional recommender systems~\cite{10.1145/3523227.3546767, kim2024large, gao2023chatrec,liu2023chatgpt}. Specifically, in system-centric evaluation~\cite{jadidinejad_simpsons_2021}, users' rating history are used in the prompt for LLMs, and if the recommended items from this LLM-based RS match a user's preference, it is treated as a good recommendation, and traditional metrics (e.g. MAE~\cite{10.1145/3523227.3546767}, RMSE~\cite{10.1145/3523227.3546767}, NDCG~\cite{liu2023chatgpt}, Hit Ratio~\cite{di_palma_evaluating_2023}, Recall~\cite{di_palma_evaluating_2023}, Precision~\cite{di_palma_evaluating_2023} etc) are deployed accordingly. 






However, LLMs are pre-trained models on extensive textual data drawn from various sources, including articles, books, websites, and other publicly accessible written materials and trained on billion parameters~\cite{lin2024dataefficient}. 
Thus, LLMs exhibit black-box and probabilistic characteristics, meaning their internal working is not visible, which results in different outputs for the same input across different iterations. 
Moreover, the evaluation becomes extremely challenging when the correct output is not known, such as in the case of top-$k$ recommendations. This leads to the emergence of test oracle problem in evaluating LLM-based recommender systems. The test oracle problem refers to a challenging problem where validating if the computed output is correct or not during the testing of data-intensive softwares, as most of the times the output is not known. 
So, this leads to issues in the quality of outputs and a thorough evaluation is much needed for these systems: using an LLM such as GPT for recommender systems requires more than just the evaluation of generated output (ratings or recommendations), as typically done in traditional recommender systems. 

Metamorphic Testing (MT) ~\cite{chen_metamorphic_2019} is introduced to handle the test oracle problem in the field of software testing. 
MT is a software testing technique based on the defined generic relations, Metamorphic Relations, between inputs rather than conventional mapping the input with the output~\cite{mao_empirical_2024}. The input is generated using any test case generation strategy and a follow up test input is generated using the defined metamorphic relations (MRs). Both the inputs are tested and the generated outputs are compared. If the defined relation exists
in the outputs then the testing is said to be successful~\cite{chen_metamorphic_2019, mao_empirical_2024}. 
While recent studies show the usage of Metamorphic Testing for the evaluation of LLMs~\cite{hyun_metal_2023}, chatbots~\cite{bozic_testing_2019}, and traditional RS~\cite{mao_empirical_2024}, there is a gap about the evaluation of MT in LLMs-based RS, which is the focus of this study. 


This paper thoroughly evaluates the performance of GPT-based recommender systems using the metamorphic testing techniques from both RS and LLMs perspectives. Specifically, we evaluated four MRs relations: rating multiplication, rating shifting in RS; and adding spaces, randomness in the prompt for ChatGPT. We introduced a framework to control the randomness in the outputs from GPT-based RS  so as to form a solid and consistent basis to evaluate the MRs. Both Kendall $\tau$ and Ranking Biased Overlap (RBO) are selected to measure whether the relationship in MRs are satisfied in their corresponding outputs, so that both the complete ranked recommendation list and their ranking positions are took into consideration. The contributions of this paper are: 
\begin{itemize}
    \item To the best of our knowledge, this is the first work proposing Metamorphic Evaluation for LLMs-based RS. 
    \item A framework is proposed to control the randomness in the outputs from GPT-based RS. 
    \item Results from MRs in both RS and LLMs indicate that there is need to evaluate LLMs-based RS differently from traditional RS. 
\end{itemize}


\section{Related Papers}
\subsection{LLM-based Recommender Systems}
 

Due to LLMs\text{'} advanced capabilities, LLMs have the potential to significantly transform the field of recommender systems.
The use of language models in recommender systems, such as LMRecSys~\cite{Zhang2021}, generally utilises prompt generation by providing the user history for few-shot recommendations. The prompt tuning are of two types - continuous vector embeddings as prompts~\cite{gu2022ppt} and discrete prompts using text tokens~\cite{gao-etal-2021-making}. For example, the “Pretrain, Personalized Prompt, and Predict Paradigm” (P5)~\cite{10.1145/3523227.3546767} combines recommendation tasks and uses instruction-based prompt design with detailed description in a natural language format. Several works indicate that the instruction based prompts are promising for NLP-related tasks as they are flexible and similar to humans communication~\cite{efrat2020turking}. The research on LLM-based RS majorly uses instruction-based recommendation for different LLMs such as Alpaca ~\cite{10.1145/3604915.3610647}, GPT ~\cite{liu2023chatgpt}, Google Palm ~\cite{di_palma_evaluating_2023} which assisted in addressing the sparse user-item matrix problem.

\subsection{Evaluation Framework in RS}

For the evaluation of traditional RS, three primary evaluation framework approaches are commonly used: online evaluation, offline evaluation, and user-based evaluation \cite{jannach_survey_2021}. Various metrics have been proposed to check the reliability and robustness of the systems.  Some of the commonly used metrics for the evaluation of RS are recall, precision, f1-score, mean absolute error (MAE), root mean squared error (RMSE) \cite{10.1145/963770.963772}. To assess the scoring and ranking of a list of items, normalized discounted cumulative gain (nDCG) \cite{10.1145/582415.582418} and mean reciprocal rank (MRR) \cite{jannach_recommender_2021} are the commonly employed metrics. Currently, for the evaluation of LLM based RS, the researchers have been following the same evaluation metrics as of traditional RS, and since LLM-based RS is at its early stage, the evaluation primarily consists of offline evaluation.

\subsection{Metamorphic Testing}

The evaluation of software systems involves choosing input samples for the system execution and then comparing actual outputs with expected results to detect any failures~\cite{hyun_metal_2023, chen_metamorphic_2019}. Recommender systems evaluation works in the similar way: during system-centric evaluation, the dataset is divided into training and test data; the model is trained on the training data and test data is used to evaluate the system.
But during the implementation it faces a primary challenge - test oracle problem which refers to the case where validating if the output given by the software is the desired output for any given input~\cite{chen_metamorphic_2019}. To overcome this problem, a property-based software testing technique, metamorphic testing (MT) was introduced by Mao et. al.~\cite{mao_empirical_2024} for the evaluation of traditional recommender systems. MT works on defining generic relations (metamorphic relations (MR)) between inputs and outputs and look for violations of these relations during testing. In addition, Hyun et al.~\cite{hyun_metal_2023} proposed a metamorphic testing framework (METAL) for LLMs as a language model. Josip Bozic and Franz Wotawa~\cite{bozic_testing_2019} tested the working of chatbots using metamorphic relations by introducing unexpected user responses during the conversation. 

\subsection{Gaps}

Existing research mainly works on the MT for traditional RS (e.g. Collaborative Filtering)~\cite{mao_empirical_2024}, LLMs tasks~\cite{hyun_metal_2023} and on conversational chatbots~\cite{bozic_testing_2019}. This study focus on the gap of MT for LLMs-based Recommender Systems from both the recommender system perspective and the LLMs perspective to their recommendation capability. To achieve this, we developed a framework to control the randomness introduced by LLMs's probabilistic characteristics so as to examine the MRs in this context. 


\section{Methodology}








\subsection{Overview}

\begin{figure}[htbp]
    \centering    \includegraphics[width=0.5\textwidth]{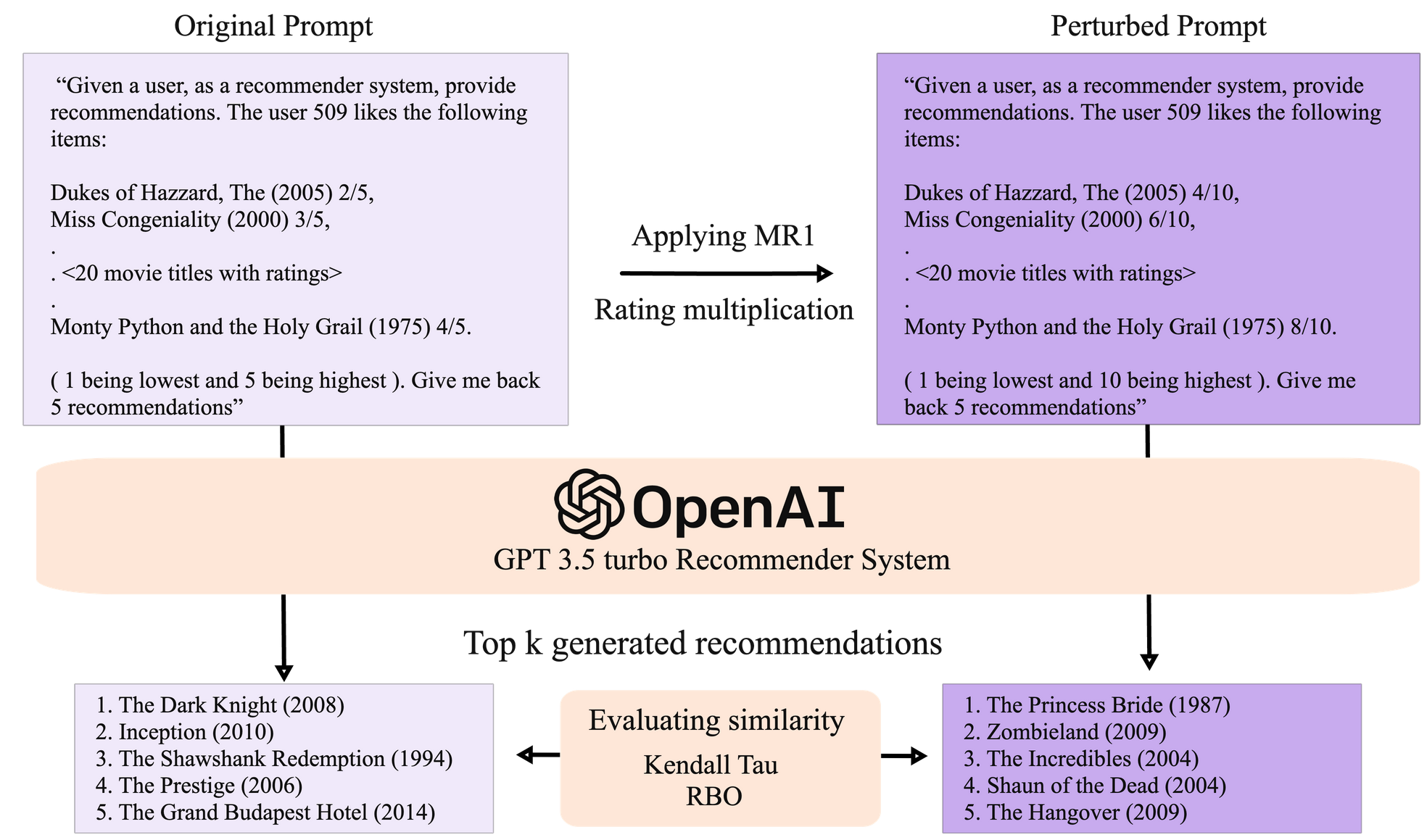}
    \caption{Overview with MR1 as an example}
    \label{fig:overview}
\vspace{-0.4cm}
\end{figure}

\begin{table*}[htbp]
\centering
\caption{Example of Prompt Perturbations} \label{table:exampleMRs}
{\scriptsize
\begin{tabular}{ p{2.5cm} | p{14.5cm} }
  \toprule
  \textbf{Method} & \textbf{Prompt} \\ [0.5ex]
  \midrule
  Original Prompt & ``Given a user, as a recommender system, provide recommendations. The user 509 likes the following items: Dukes of Hazzard, The (2005) 2/5, Miss Congeniality (2000) 3/5, Click (2006) 1/5, Ultraviolet (2006) 2/5, Monty Python and the Holy Grail (1975) 4/5. (1 being lowest and 5 being highest ). Give me back 5 recommendations'' \\ [0.5ex]
  \hline
  MR1: Rating multiplication  & ``Given a user, as a recommender system, provide recommendations. The user 509 likes the following items: Dukes of Hazzard, The (2005) 4/10, Miss Congeniality (2000) 6/10, Click (2006) 2/10, Ultraviolet (2006) 4/10, Monty Python and the Holy Grail (1975) 8/10. (1 being lowest and 10 being highest ). Give me back 5 recommendations'' \\ [0.5ex]
  \hline
  MR2: Rating shifting & ``Given a user, as a recommender system, provide recommendations. The user 509 likes the following items: Dukes of Hazzard, The (2005) 3/6, Miss Congeniality (2000) 4/6, Click (2006) 2/6, Ultraviolet (2006) 3/6, Monty Python and the Holy Grail (1975) 5/6. (2 being lowest and 6 being highest ). Give me back 5 recommendations'' \\ [0.5ex]
  \hline
  MR3: Adding spaces &  ``Gi ven a u ser , as a re co m mende r syst em, prov ide r e commend ati ons. T he use r 509 l ik e s t he follow i n g item s : Dukes of Ha zza rd, T h e (2005 ) 2/5, M i ss Cong e n ialit y ( 2 0 00) 3/ 5, Cli ck (2 006) 1/5 , Ul t r aviolet ( 20 06) 2 /5 , Monty P y th o n and th e Ho ly Grail (1 975) 4/ 5. ( 1 be ing lowest and 5 be i ng hi ghes t ). Giv e me back 5 recommen d a ti o ns.''  \\ [0.5ex]
  \hline
  MR4: Adding random words &   ``Given a user, as a banana recommender system, grape provide recommendations. The user pear 509 likes banana the following items: grape Dukes of banana Hazzard, The (2005) 2/5, Miss Congeniality (2000) pear 3/5, Click (2006) 1/5, Ultraviolet (2006) 2/5, Monty Python and the Holy Grail (1975) 4/5. (1 being lowest and 5 being grape highest ). Give me back 5 banana recommendations, banana one movie per line and don't give any explanation'' \\ [0.5ex]
 \bottomrule
\end{tabular}
}
\vspace{-0.4cm}
\end{table*}

The metamorphic testing operates by generating source test inputs by using any test case generation strategies, and MRs are defined for these inputs on the basis of the properties of the system. The follow up test inputs are generated on the basis of the defined MRs and then the outputs for both the inputs are calculated and evaluated. If the defined MR is maintained between the outputs, the program is said to succeed. We leverage the same testing strategy to check for the robustness of the GPT-based RS, and the overview of the proposed Metamorphic Testing framework for GPT-based RS is shown in Figure~\ref{fig:overview}, 
which showing the overall process of one MR relation in GPT-based RS. 
The proposed methodology includes three main components: 
i) Prompt Construction; ii) Metamorphic Relations; iii) Output Refinement for randomness control. 


\subsection{Prompt construction}

Here, we used the instruction based few-shot discrete prompts which is close to human language. The prompt was based on ``person pattern" prompt which was proposed by White et. al.~\cite{white2023prompt}. The prompt comprises both static and variable components. The static portion specifies the action GPT is tasked with, while the variable part provides the context or parameters for executing that action ~\cite{di_palma_evaluating_2023}. A generic example of the prompt is given below: 

Prompt - \textit{``Given a user, as a recommender system, provide recommendations. The user \{user\} likes the following items: \{movies\}. (1 being lowest and 5 being highest).Give me back 5 recommendations''}

An example of the Original Prompt and the MRs in this study are shown in Table~\ref{table:exampleMRs}.


\subsection{Metamorphic Relations}

\subsubsection{MT of the \textbf{Ratings} in LLM Based RS}
To evaluate the top-k generated recommendations, we considered different ratings that represent the same preference of the user. This involves applying prompt perturbation on just the ratings part of the prompt, keeping rest exactly the same as the original to check if the model is sensitive to the ratings scale and generated recommendations are consistent that represents user’s preferences.  

To observe the impact of changing of the rating scale on the top-$k$ generated items, following~\cite{mao_empirical_2024}, we defined two metamorphic relations in ratings from RS that were used for the evaluation:

\textbf{$\bullet$ MR1: Rating multiplication} - For all the items in the prompt for each user, the ratings for each item and the total rating is multiplied by a constant $\lambda$ integer, e.g. that original ratings, $R/5$ becomes $\lambda[R/5]$.

\textbf{$\bullet$ MR2: Rating shifting} - For all the items in the prompt for each user, the ratings for each item and the total rating is shifted, either increased or decreased, by a constant $\lambda$ integer such that original ratings, $R/5$ becomes $(\lambda + R)/(\lambda + 5)$. 

\subsubsection{MT of \textbf{Prompts} in LLM based RS}
To evaluate the robustness of the LLM, we do metamorphic testing on the language part of the prompt to check the impact on the performance.  This involves applying linguistic variations on the prompt to see how language variations can affect’s model performance. By comparing the responses to these paraphrased inputs, we can assess whether the LLM consistently produces accurate and relevant answers despite the changes in semantic structure of the prompt. 

Following~\cite{hyun_metal_2023}, we defined two metamorphic relations in language from LLMs perspective that were used for the evaluation. 
We have used  semantic-preserving prompt perturbation for the linguistic manipulation of the prompt. This evaluation helps uncover potential weaknesses in the model's understanding and processing of differently formatted input using diverse linguistic contexts.
    
\textbf{$\bullet$ MR3: Adding spaces}  - This relation works by inserting spaces between characters in the given prompt containing user history and rating.

\textbf{$\bullet$ MR4: Adding random} - This relation works by inserting random words in the given prompt containing user history and rating, such as ``apple'', ``grape'', ``banana'', ``pear''.

\subsection{Output Refinement for Randomness Control}\label{sec-control}
ChatGPT introduces an element of randomness while generating outputs to incorporate diversity. Namely, this randomness can lead to varied results for the generated recommendations during each iteration~\cite{liu2023chatgpt}. This creates difficulty in evaluating the GPT-based RS if output is very different the each time. To handle this, we controlled the randomness of the generated output through prompt engineering by considering the following two variables: 
\begin{itemize}
    \item $l$: the number of items provided in the prompt.
    \item $k$: the number of items in top-$k$ recommendation list. 
\end{itemize}
To check for the impact on these variables on the recommendations and randomness during different iterations, similarity metrics were used - Kendall $\tau$, Ranking Biased Overlap and overlap between the lists. 
Finally, 
$l$ and $k$ are determined when consistent outputs are obtained in different iterations in terms of the above three similarity metrics. 
Detailed experiment results about this is in Section~\ref{sec-exp}. 



\section{Experiment}\label{sec-exp}

\subsection{Experiment Configuration}

\textbf{Dataset}: For the experiments we used the MovieLens 100k dataset by Grouplens\cite{GroupLens_2021}. The dataset contains 100,000 ratings and 3,600 tag applications applied to 9,000 movies by 610 users. 
\textbf{GPT-based RS}: The recommendations are generated using GPT 3.5 \textit{turbo} model. 
\textbf{Evaluation of MTs}: The relationship between input and outputs of MTs are measured with the following similarity metrics: Kendall $\tau$~\cite{jadidinejad_simpsons_2021}, Ranking Biased Overlap (RBO)~\cite{10.1145/1852102.1852106} and overlap ratio between the top-$k$ recommendation lists. The $t$-test with a 95\% confidence level has been applied to evaluate whether the results is statistically significant.



\subsection{Results of Randomness Control}

As discussed in Sec~\ref{sec-control}, we control the randomness of the generated outputs from GPT-based RS with two variables: $l$ the number of items provided in the prompt, and $k$ the number of items in top-$k$ recommendation list. 

The results for $k$ is shown in Table~\ref{tab:recommendations}. 
In the prompt construction here, we used all movies in user history if their corresponding ratings are greater than 3, which indicates a positive preference. 
We conducted two iterations for each user and evaluated similarity by comparing the recommendation lists for each user from these two iterations. 
It is observed that the bigger the $k$, the more randomness in the generated recommendation list. Specifically, the top 5 recommendations perform the best. 
The higher RBO scores suggest that the recommendations at the top of the list are more consistent across iterations compared to that of larger $k$, as RBO assigns greater weight to top-ranked items while Kendall $\tau$ assesses the overall order of items in the list. So, to ensure consistency in metamorphic testing, we conduct experiments focusing on the top 5 recommendations.


\begin{table}[htbp]
\centering
\caption{Results of $k$ in different top-$k$ recommendations}
\label{tab:recommendations}
\begin{tabular}{@{}lcccc@{}}
\toprule
\textbf{$k$} & \textbf{Kendall $\tau$} & \textbf{RBO} & \textbf{Overlap Ratio} &  \\ \midrule
$k=5$ & 0.8784	& 0.9632 & 0.8146 \\
$k=10$ & 0.8673	& 0.951 & 0.9445 \\
$k=30$ & 0.7679	& 0.9068 & 0.8801 \\
$k=50$ & 0.7096	& 0.8522 & 0.6870 \\ \bottomrule
\end{tabular}
\vspace{-0.2cm}
\end{table}



The results for $l$ is shown in Table~\ref{tab:l} with various $l$ values, ranging from 5 to 30.  
Specifically, we ran 10 iterations for $l$ value and calculated the average Kendall $\tau$, RBO and overlap ratio. 
It is observed that the lists generated from different $l$ are similar as indicated by higher similarity. We opted for a selection of $l=20$ in the prompt for the proposed metamorphic testing since it represents a broader range of user interests while requiring less computational time.

\begin{table}[htbp]
\centering
\caption{Results of
$l$ the number of items per user in the prompt for top-5 recommendations}
\label{tab:l}
\begin{tabular}{@{}lccc@{}}
\toprule
\textbf{$l$} & \textbf{Kendall $\tau$ } & \textbf{RBO } & \textbf{Overlap Ratio} \\ \midrule
$l=5$ & 0.9120 & 0.9707 & 0.8495 \\
$l=10$ & 0.8757 & 0.9663 & 0.8088\\
$l=20$ & 0.9116 & 0.9768 & 0.9623 \\
$l=30$ & 0.9179 & 0.9791 & 0.9665\\ 
\bottomrule
\end{tabular}
\vspace{-0.2cm}
\end{table}




\subsection{Results of MRs}

Following the above experiments about controlling randomness in GPT-based RS, we set $k = 5$ and $l=20$ to examine the metamorphic relations: MR1, MR2, MR3 and MR4. 
Specifically, 
we ran 10 iterations of each defined MRs. The generated lists were compared against a baseline list consisting of one iteration of top-5 recommendations generated using 20 movies with no modifications. This comparison served as the reference point to assess the effectiveness of various modifications applied during the testing process. 

\begin{table}[h!]
\centering
\caption{Results of MRs}\label{tab:comparisonMRs}
{\footnotesize
\begin{tabular}{lcccc}
\toprule
\textbf{Method} & \textbf{Kendall $\tau$ (SD)} & \textbf{RBO (SD)} & \textbf{Overlap ratio (SD)} &  \\
\midrule
No change (baseline) & 0.9116 (0.0055) & 0.9768 (0.0027) & 0.9623 (0.0030) \\
MR1: Multiply & 0.4829 (0.0082) & 0.8496 (0.0021) & 0.8146 (0.0025)\\
MR2: Addition & 0.4966 (0.0057) & 0.8460 (0.0014) & 0.8174 (0.0028)\\
MR3: Spaces & 0.0640 (0.0121) & 0.4710 (0.0081) & 0.4882 (0.0057)\\
MR4: Random words & 0.2295 (0.0117) & 0.6802 (0.0050) & 0.6863 (0.0089)\\
\bottomrule
\end{tabular}
}
\end{table}

Table~\ref{tab:comparisonMRs} shows the results of evaluating different Metamorphic Relations (MRs) (averaged over 10 runs with standard deviation (SD)), and the unpaired $t$-test with 95\% confidence level is deployed to examine whether the difference between each MR output and the baseline list: the corresponding $p$-values are $< 0.0001$, which means they are all statistically significantly different to the baseline list.

$\bullet$ \textbf{MR1}: The MR1 involves multiplying the ratings by a constant $\lambda$ while representing the same user preferences. After introducing this MR to the prompt, it can be seen that the average Kendall $\tau$ value dropped significantly, suggesting a reduced overlap between the lists. Interestingly, the average RBO value is still over 0.8, indicating that despite modifications the top items of the generated lists for each user are still similar as compared to the bottom. The small standard deviation indicates that there is a consistency in the results. 
    
$\bullet$ \textbf{MR2}: Similar to MR1, this relation involves adding a constant $\lambda$ in all the ratings, thereby changing the rating scale while keeping the same user preferences. The Kendall score is slightly higher than the average Kendall score of MR1 but the similarity in the order of the items generated is still poor. The RBO value reveals a similar trend, representing similarity among the top items of the list.
    
$\bullet$ \textbf{MR3}: This MR is introduced to the prompt to check for the performance under language variation. After introducing spaces in the prompt, the average Kendall and RBO values dropped very low indicating minimal correlation and overlap between the lists. The small standard deviations also show an inconstancy in the generated output.

$\bullet$ \textbf{MR4}: In this MR, random words are added in the prompt to check for performance. MR4 performs better than MR3 with higher average Kendall and RBO values but overall the similarity is low when compared with MR1 and MR2 which means that the generated lists are very different. 

\subsection{Discussion}



It can be observed that MR1 and MR2 performed better than MR3 and MR4. This suggests that changes in the rating scale had a relatively more predictable and consistent impact on the recommendation outcomes. In contrast, altering the semantic structure of the prompt led to less desirable results, indicating a greater degree of variability or unpredictability in the system's response. Moreover, while the average Kendall $\tau$ values showed a significant drop across all Metamorphic Relations (MRs), the average values for Rank-Biased Overlap (RBO) didn't decrease as drastically. This suggests that despite changing the prompts after the application of MRs, the top items in the generated list remained similar across users. However, as we move down to the bottom of the list, the order of the recommendations changes, as indicated by average Kendall's $\tau$. In addition, there are certain limitation to the experiments which includes internal validity. We repeated the experiments 10 times for each results to get an overall performance of the GPT-based RS under different situations. 

\section{Conclusion}


This paper proposes metamorphic testing for LLM-based RS using metamorphic relations. Our experiments conducted on the GPT 3.5 using the MovieLens dataset revealed insights into how these metamorphic relations influence the recommendation lists generated by the system. Our findings revealed noticeable decrease in similarity in the generated outputs using multiple MRs and prompt variations. 
This exploration serves as an initial step towards integrating metamorphic testing into the evaluation framework for LLM-based RS. Moving forward we intend to test multiple MRs in our future work to gain comprehensive understanding of the system. 

\bibliographystyle{IEEEtran}
\bibliography{sample_sigconf_biblatex}



\end{document}